\documentclass[pra,aps,10pt, twocolumn, floatfix]{revtex4-1} 

\usepackage {graphicx} 
\usepackage {amssymb} 
\usepackage {amsmath}

\newcommand{\reffig}[1]{Fig.~\ref{#1}}
\newcommand{\refeq}[1]{Eq.~(\ref{#1})}

\begin{document}

\title{Self-induced transparency mode-locking, and area theorem}

\author{R. M. Arkhipov}

\affiliation{ITMO University, Kronverkskiy prospekt 49, 197101, St. Petersburg,  Russia}

\author{M.V. Arkhipov}

\affiliation{Faculty of Physics, St. Petersburg State University, Ulyanovskaya 1, Petrodvoretz, St. Petersburg 198504, Russia}

\author{I. Babushkin}

\affiliation{Institute of Quantum Optics, Leibniz University Hannover,
  Welfengarten 1 30167, Hannover, Germany}
\affiliation{Max Born Institute, Max Born Str. 2a, Berlin, Germany}

\author{N.N. Rosanov}

\affiliation{Vavilov Optical State Institute, St.Petersburg, Russia}
  




\begin{abstract}
  Self-induced transparency  mode-locking 
  (or coherent mode-locking, CML) which is based on intracavity self-induced transparency
  soliton dynamics, allows potentially to achieve nearly single cycle
  intracavity pulse durations, much below the phase relaxation time $T_2$ in
  a laser, which, despite of great promise, has not yet been realized
  experimentally. We develop a diagram technique which allows
    to predict the main features of CML regimes in a
    generic two-section laser. We show that CML 
    can arise directly at the first laser threshold if the phase
    relaxation time is large enough. Furthermore, CML regimes can be unconditionally
  stable.   We also predict
  the existence of
  ``super-CML regimes``, with a pulse coupled to several Rabi oscillations
  in the nonlinear medium.
\end{abstract}

\pacs{}

\maketitle

Ultrashort optical pulses with durations of a few hundreds 
femtoseconds (or picoseconds) are used in different applications
ranging from real-time monitoring of chemical reactions to
high-bit-rate optical communications
\citep{keller2010ultrafast}. The main technique 
to obtain such pulses nowadays is a passive
mode-locking. The pulse duration
in such lasers is fundamentally limited by
the
inverse bandwidth of the gain medium $\gamma_2=1/T_{2}$.

This very basic limitation is not valid anymore when the electric field
is so strong, that Rabi oscillations in both gain and absorber sections are
existed.  The general idea of coherent modelocking (CML)
  \citep{kozlov1997self,kozlov1999self,menyuk2009self,talukder2009analytical,
  kozlov2011obtaining,kozlov2013single,arkhipov2015coherent,arkhipov2014selfa}
  is that a self-induced transparency (SIT) soliton (or $2\pi$ pulse)
  \cite{mccall1969,mccall1969self,kryukov70,Allen:book,lamb1971analytical}
  propagates stably in the absorber, whereas the gain section is
  arranged in such a way that the very same pulse (with the same
  shape) propagates there as a $\pi$ pulse. Because of this, it
  ``feeds'' the energy from the gain medium
  \cite{mccall1969self,kryukov70,Allen:book,lamb1971analytical}, which
  makes it shorter. At the end, a stable equilibrium is set up due to
  balance of losses and gain, which defines finally the pulse
  duration. If the linear losses are small enough, the pulse duration
  will decrease until stopped by other mechanisms such as intracavity
  dispersion \cite{vysotina09}. 

Up to now there were no clear
experimental observation of the CML regime, although some work was
made to improve theoretical description and to observe  some elements of
CML experimentally. At the same time, in
\cite{menyuk2009self,talukder2009analytical} the CML was predicted
theoretically to quantum cascade lasers; 
in
Ref.~\citep{arkhipov2015mode}  experimental evidence of
mode-locking regime in a laser with a coherent absorber has
  been reported
(the gain section remained in incoherent regime). 

In the present work, we show that CML might be easier accessible than
was though before. We develop a technique based on the well known
McCall and Hahn area theorem which determines the evolution of the
pulse area in the CML laser. Using this technique we show that in a
wide range of parameters CML regime is a stable attractor of the
dynamics.  We demonstrate furthermore the regimes we call super-CML
where pulse area in absorber may approach $2n\pi$ for $n=0$ and
$n>1$ instead of the only known case of $n=1$. In order to keep the
  theory simple we will use the parameters where the pulse duration is
  far above the single-cycle pulse duration but still
  in the femtosecond range.
 
\begin{figure*}[tpbh]
\center{\includegraphics[width=0.9\linewidth]{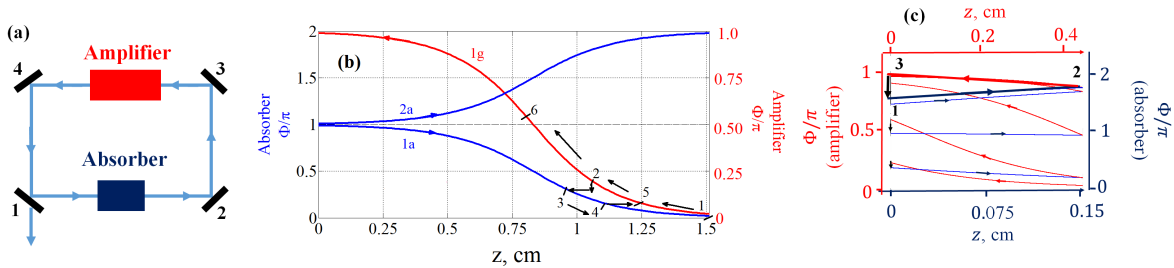}}
\caption{ (a): Schematic representation of mode-locked laser with a ring
  cavity where unidirectional counter clockwise lasing is assumed, $1,2,3,4$ are cavity mirrors,(b): Branches of solution of Eq.(\ref{eqAreatheorem}) for the
  absorber 1a, 2a (blue lines) and the amplifier 1g, 2g (red
  line) for $|\alpha_0| = 10$ cm$^{-1}$,  and (c): evolution of the pulse area $\Phi$ in the absorber and gain
  medium  from the initial value $\Phi_0=0.025\pi$ to a limit cycle
  (thick arrows $123$).}
\label{fig:McallHahngr1}
\end{figure*}
An important quantity describing the pulse dynamics in the coherent
regime is the pulse area, defined as \citep{mccall1969self}
$\Phi(t,z)=\frac{d}{\hbar} \int_{-\infty}^{t} \mathcal{E}(t',z) dt'$,
%
where $d$ is the transition dipole moment of two-level particles,
and $\mathcal{E}(t,z)$ is the pulse envelope. The area $\Phi$ of a
stable
SIT-induced soliton  is $2\pi$.
The basic result of the SIT theory is the
so-called pulse area theorem. This theorem governs the evolution of
the pulse area $\Phi$ during its propagation in a two-level absorbing
(or amplifying) medium with an inhomogeneously broadened line
\citep{mccall1969self,kryukov70,Allen:book}:
\begin{equation}
\frac{d\Phi}{dz}=-\frac{\alpha_0}{2} \sin\Phi,
\label{eqAreatheorem}
\end{equation}
where
$\alpha_0 = \frac{8\pi^2 N_0 d^2 \omega_0 T^{*}_{2}}{\hbar c}$
is the absorption coefficient per unit length, $N_0$ is the
concentration of two-level particles, $\omega_0$ is the medium
transition frequency and $\gamma^{*}_2=1/T^{*}_{2}$ is the width of inhomogeneously
broadened line. We remark that \refeq{eqAreatheorem} is valid
  in the limit of negligible homogeneous broadening $T_2\to\infty$. The solution of \refeq{eqAreatheorem} is
$\tan(\Phi/2)=\tan(\Phi_0/2)e^{-\alpha_0 z/2}$,
where $\Phi_0$ is the initial pulse area.$\Phi=\pi$ is an
  unstable steady-state of \refeq{eqAreatheorem} whereas $\Phi=0,2\pi$
  are stable ones.
Coherent pulse propagation in an amplifying medium can be also
described by \refeq{eqAreatheorem} simply assuming the opposite sign of
$\alpha_0$.  From \refeq{eqAreatheorem} it follows that the same
diagram can be used, only the opposite $z$-direction is to be taken.
In this case, $\Phi=\pi$ is a stable point.
Thus, a stable $2\pi$ soliton in an absorbing
  medium, and a stable $\pi$ soliton in a gain medium can be formed.

\begin{figure*}[tpbh]
\center{\includegraphics[width=1.\linewidth]{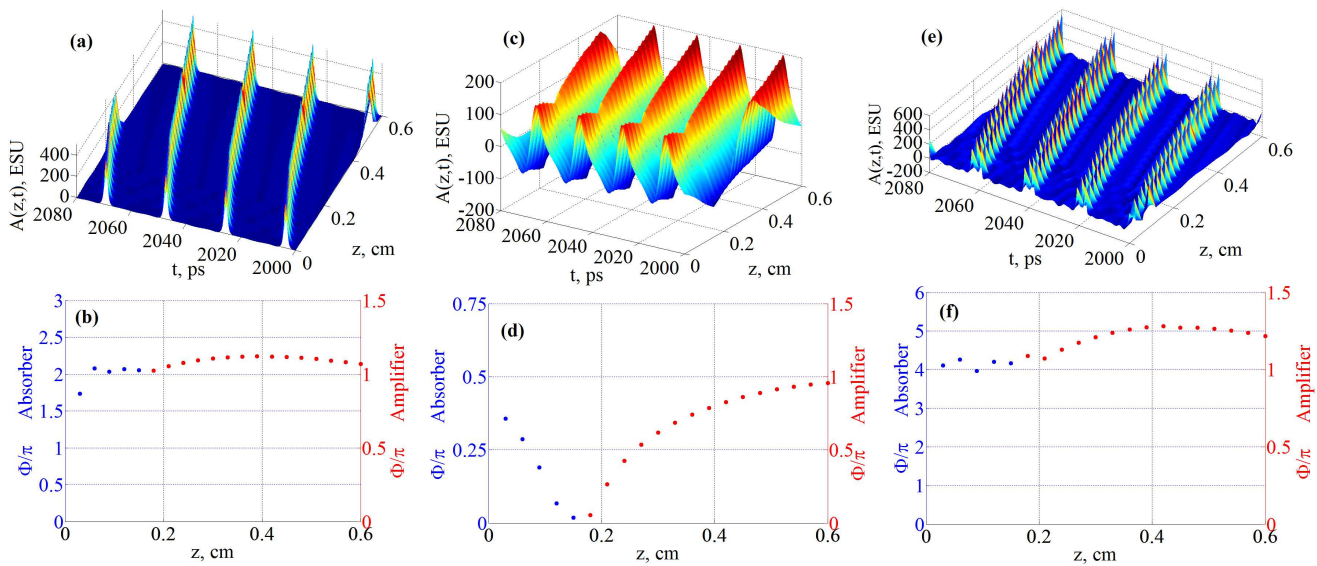}}
\caption{(a),(c),(e): Distribution of the electric field amplitude $A(z)$
   in the cavity for different values of $m_d$: (a): $m_d=2$, (c):
  $m_d=0.5$, (e): $m_d$=4. (b),(d),(f): Evolution of the pulse area in the cavity
  during one round-trip for different values of $m_d$: (b): $m_d=2$,
  (d): $m_d=0.5$, (f): $m_d$=4. The other parameters are the
  following: Central wave length $\lambda_{a,g}=0.7 \mu m$, reflectivity of the mirror $r=0.8$, length
  of the medium $L_g=0.45$ cm, $L_a=0.15$ cm, $d_g=5$ Debye, $N_{0g}=N_{0a}=12.5\cdot  10^{14}$ cm${^{-3}}$, $T_{1g}=T_{1a}=0.16$ ns,
  $T_{2g}=T_{2a}=40$ ps.  High values of $T_2$ (units and hundreds of ps) can be obtained in quantum dots at low temperatures \cite{borri2001ultralong}.}
\label{fig:McallHahngr4}
\end{figure*}

Consider a CML laser operating in a unidirectional
lasing regime shown in \reffig{fig:McallHahngr1}a with
 the absorbing and amplifying media 
separated in space. In order to achieve
  the same pulse shape for $2\pi$-pulses and $\pi$-pulses in the absorber and amplifier
  respectively, one should take $d_a=2d_g$, where $d_{a(g)}$
  is the dipole moment of the absorber (amplifier) \citep{kozlov1997self,kozlov1999self,menyuk2009self,talukder2009analytical,
    kozlov2011obtaining,kozlov2013single,
    arkhipov2015coherent,arkhipov2014selfa}.
 The branches of solutions of Eq.~(\ref{eqAreatheorem})
in the amplifier and absorber are shown in
Fig.~\ref{fig:McallHahngr1}b. Because of different dipole
  moments, these two branches
 are also
different. Besides, as said, we 
reverse the sign of $z$ to ``mimic'' the change of sign of $\alpha$
in \refeq{eqAreatheorem}. We
choose the positive direction
of motion in $z$ as the pulse propagates in  the absorber (blue curves noted as $1a$,
$2a$), whereas for the amplifier
the propagation direction is opposite (red curve noted as $1g$).

Using such diagram technique we are able to 
understand easily the details of the pulse area evolution.
In our laser in
  \reffig{fig:McallHahngr1}a a short pulse
 passes
through the amplifier, is reflected from the mirror $1$ with the
reflection coefficient $r$, and then enters
the absorber. We assume for simplicity that the
other mirrors do not produce any energy loss.  We also assume that the
pulse travels long enough in every cavity section such that both the
gain and absorber are able to recover to their equilibrium values
between the pulse passages.  Using the diagram in \reffig{fig:McallHahngr1}b
we are able to follow the evolution of the
pulse area during one round-trip in the ring laser. 

Let us now suppose
that a short pulse with the small initial area $\Phi_0 \ll \pi$ enters
the gain section (point $1$ on the red (amplifier)
curve $1g$).  As the pulse propagates in the amplifier , the
corresponding point on the diagram is moving from right to left along
the amplifier branch to the point $2$, which
is accompanied by an increase of the pulse area as one can see
  from \reffig{fig:McallHahngr1}b. After the pulse
passes the amplifier, it is reflected by a non-ideal mirror and its
area is thus reduced,
  which corresponds to the moving the point on the diagram \reffig{fig:McallHahngr1}b
   vertically
from $2$ by some amount depending on $r$. Then, by moving along the straight line
parallel to the horizontal axis one places the
point to $3$ on the absorber branch, which
now describes the pulse entering the absorber. Evolution of the pulse
in the absorber is described by moving from left to right along the
blue branch until the point $4$ where the pulse leaves the absorber. Now this point returns back to the
gain branch at the point $5$ thus closing the
  propagation circle. It continues to move further to the point $6$ etc.  
Using such an approach one can expect that after many round-trips a
stable limit cycle appears.  This cycle and its formation is
presented in Fig.~\ref{fig:McallHahngr1}c.

To confirm the results of our qualitative analysis presented above  we perform
 numerical simulations based on a set of Maxwell-Bloch equations
describing propagation of light in a two-level medium in a
slow-varying envelope approximation \citep{kryukov70,Allen:book,lamb1971analytical,arkhipov2015coherent,arkhipov2014selfa}:
\begin{gather}
\partial_tp_s(z,t)=-\gamma_2(z)p_s(z,t) +
g(z)n(z,t)A(z,t),\label{eq:eqPsChap2} \\
\partial_tn(z,t)=-\gamma_1\left[n(z,t)-N_0(z)\right]-F(z,t), \label{eq:eqPhoChap2}\\ 
\partial_t A(z,t) + c\partial_zA(z,t)=
\kappa(z) p_{s}(z,t), \label{eq:eqAcChap2}
\end{gather}
where $g(z)=\frac{d(z)}{2\hbar}$, $\kappa(z)=-4\pi\omega_0(z)
d(z)N_0(z)$, 
$F(z,t) = 4g(z)A(z,t)p_s(z,t)$, 
$p_s(z,t)$ is the slowly varying envelope of imaginary part of non-diagonal element of the
quantum mechanical density matrix of a two level
  particle, $n(z,t)$ is the
population difference between the
lower and upper energy levels, $A(z,t)$ is the
slowly varying amplitude of the electric field.  The equations include parameters of the two-level
particles, such as transition dipole moment $d(z)$, concentration of
two-level particles $N_0(z)$, population difference relaxation
time $T_{1}(z)=1/\gamma_1(z)$, polarization relaxation time
$T_{2}(z)=1/\gamma_2(z)$ and transition frequency of the
two-level medium $\omega_0(z)$. We also assume that $g(z)$ is a constant in every piece of
the cavity defined, namely $g(z)\equiv
g_a$ in the absorption medium and $g(z)\equiv
g_g$ in the gain medium (and analogously for $\kappa(z)$, $d(z)$,
$N_0(z)$, $\omega_0(z)$, each of them obtains an 
index $a$ or $g$ in the corresponding medium). In
particular, $N_{0g}=-1$, $N_{0a}=1$.
 The set of equations
(\ref{eq:eqPsChap2})-(\ref{eq:eqAcChap2}) allows accurate modeling of evolution of
extended two-level media in a cavity assuming relatively long
  (up to 100 fs) pulse durations and low intensities (Rabi frequency
  $\Omega_R\ll\omega_0$), so that no significant intracavity dispersion effects and no
  multilevel dynamics enter into play. The
  equations take also into account longitudinal multi-mode
dynamics and the nonlinear coherent effects accompanying
interaction of the light with the two-level particles. We use the
system of equations (\ref{eq:eqPsChap2})-(\ref{eq:eqAcChap2}) to
analyze the  dynamics of the ring CML laser assuming the regime of unidirectional propagation
 far away from the single-cycle limit.

\begin{figure}[tpbh]
\center{\includegraphics[width=1\columnwidth]{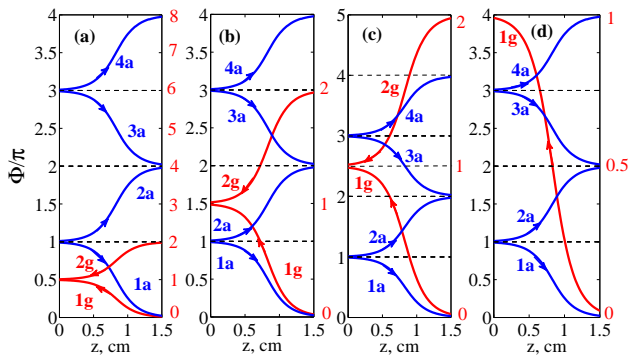}}
\caption{Branches of solution of Eq.~(\ref{eqAreatheorem}) for the
  absorber (blue lines) and amplifier (red lines) for
  different values of
  $m_d$. (a): $m_d=0.5$, (b):
  $m_d=1.5$, (c): $m_d=2.5$, (d): $m_d=4.0$}
\label{fig:McallHahngr3}
\end{figure}

Fig.~\ref{fig:McallHahngr4}a shows the distribution of the electric
field amplitude $A(z)$ in the cavity over 100 round-trips in a steady-state when absorber/gain dipole moment ratio
$m_d\equiv d_a/d_g=2$. Fig.~\ref{fig:McallHahngr3}b illustrates the dependence of
the pulse area inside the cavity during a single pass. It is clearly
seen that, as it was predicted above, the pulse area is close to
$2\pi$ in the absorber and to $\pi$ in the gain. It is remarkably that in the present
case the pulse area decreased due to the reflection from the mirror is restored in the absorber. Besides the pulse area in the
amplifier
  changes nonmonotonically.  In general however, the stable regime do
establishes with the pulse area dynamics as predicted in Fig.~\ref{fig:McallHahngr1}c.
It can be also seen in Fig.~\ref{fig:McallHahngr4}a
 that the pulse velocity in the
absorber is smaller than in the amplifier, which manifests itself in a
bend of the pulse propagation trajectory on the boundary between the
absorber and amplifier.

With the help of the diagrams introduced above we will extend
  our consideration to a more general situation when the ratio of the dipole moments is not
equal to two. First, we consider the case when $m_d < 1$. The
corresponding diagram for $m_d = 0.5$ is shown in
Fig.~\ref{fig:McallHahngr3}a. In this case the limit cycle is realized
with the branch $1g$ of the amplifier and $1a$ of the absorber. On the
amplifier branch, the pulse with the area tending to $\pi$ is formed,
whereas the absorber decreases the pulse area
almost to zero (see Fig.~\ref{fig:McallHahngr4}d). Because of
the energy conservation, pulse envelope must change its sign (see
  Fig.~\ref{fig:McallHahngr4}c), and hence,  $0\pi$ pulse is formed
  \citep{rothenberg1984observation}. 
 In the example presented in Fig.~\ref{fig:McallHahngr3}b, in
  contrast,  we set $m_d = 1.5$. In this
case, the branch $1g$ of amplifier and one of the branches  $1a$
or $2a$ of the absorber can take part in the generation. In both of these possibilities the dynamics in the amplifier 
section is similar whereas in the absorber the pulse duration can
increase (branch 2a).
The situation when $m_d$ is further increased and achieves $2
< m_d < 3$, namely, $m_d = 2.5$ is shown in
Fig.~\ref{fig:McallHahngr3}c.  One branch of the amplifier $1g$ and
one of the three branches of the absorber $1a$, $2a$ or $3a$ will be
involved in this case. The cycle formed on the branch $3a$ yields a
reduction of the pulse duration, because the pulse just before the absorber has the area $\Phi > 2\pi$.  Thus, the ratio of the dipole moments
influences the pulse duration and its dynamics in the
  CML cycle. The decrease of the pulse duration with the increase of $m_d$ was predicted theoretically in \cite{arkhipov2015coherent,arkhipov2014selfa}.

In the situation when $m_d>3$ the branch $4a$ and $1g$ can participate
in generation (Fig.~\ref{fig:McallHahngr3}d). On this branch, the
formation of  $4\pi$-pulses takes
place, which are, however, as our simulations show (see Fig.~\ref{fig:McallHahngr4}e,f), split
into two $2\pi$ pulses having typically different amplitudes and
durations. That is, two pulses  over a single
cavity round-trip arise. 
The physical reason of this splitting
can be easily understood from the fact that each  $2\pi$ "part"
causes excitation and de-excitation of the medium. Therefore,
the central part of the pulse is continuously interacting
with the particles which have returned to the ground
state and hence is gradually "eaten" \cite{poluektov1975self}.

 Note, that in numerical simulations we used parameters typical for quantum dots (QDs) (relaxations times, dipole moments etc). We
    remark that semiconductor QDs seems to be
    appropriate candidates for experimental observation of CML regime
    because of large values of dipole moments as well
    as large low-temperature relaxation times $T_2$ (in excess of 500
    ps)
    \cite{borri2001ultralong}.

The diagrams presented above allow to calculate
  some characteristics of the pulse analytically. 
 The evolution of the pulse
area at some point (for instance, just at the entrance
the gain section) $\Phi^{\mathrm{(a)}}_{n+1}=\frac{d_a}{\hbar} \int\mathcal{E}dt$  at the step $n+1$ can easily related to $\Phi^{\mathrm{(a)}}_{n+1}$
as $\Phi^{\mathrm{(a)}}_{n+1}=F\left(\Phi^{\mathrm{(a)}}_n\right)$,
  \begin{equation}
    \label{eq:1}
    F\left(\Phi\right) = 2 \mathrm{arctan}\left[e_a \mathrm{tan}\left\{ rm_d\mathrm{arctan}\left(e_g \mathrm{tan}\left\{\frac{\Phi}{2 m_d}\right\}\right)\right\}\right],
  \end{equation}
where $e_a=\exp(-\alpha_a L_a/2)$, $e_g=\exp(\alpha_g
  L_g/2)$
\cite{babushkin00a}.   This
expression allows to consider the mode-locking regime as
a fixed
 point of the mapping $F$ which is stable if $|F'|\equiv|dF/d\Phi|<1$. 

Although in general $F$ (and thus $F'$) can be calculated only numerically, analytical results can be 
obtained assuming small deviation of $e_a$, $e_g$ and $r$ from one
(as it is usually happens in a typical laser). 
In this case, we  denote $e_{a}=1-\epsilon_{a}$,
$e_{g}=1+\epsilon_{g}$, $r=1-\epsilon_{r}$ and assume
$|\epsilon_{g,a,r}|\ll 1$, $\epsilon_{g,a,r}>0$. We
  also must select the branches of $\mathrm{tan}(x)$ and
  $\mathrm{arctan}(x)$ which make $F$ continuous in the vicinity of the
  fixed point.
For instance, 
    near non-lasing point $\Phi=0$ we obtain (by taking
     derivative of \refeq{eq:1} and expanding into series in $\epsilon_{g,a,r}$): 
\begin{equation}
 |F'|=1 - \epsilon_a + \epsilon_g  -  \epsilon_r + \ldots,\label{eq:2}
\end{equation}
where  $(\ldots)$ means the higher order terms in $\epsilon_{g,a,r}$. That 
is, in the linear approximation, the expression for
  $1-|F'|$ reproduces the loss-gain balance
in the system. 

Furthermore, for a CML regime  with 
 $m_d=2$ we may assume the fixed point is close to 2$\pi$, that
 is,  $\Phi^{\mathrm{(a)}}=2\pi+\epsilon$, $\epsilon\ll1$
  which, by substitution into \refeq{eq:1}, allows us to
  obtain in the lowest order:
  $\epsilon=-2\pi\epsilon_r/(\epsilon_a+\epsilon_g+\epsilon_r)+\ldots$. This
  expression can be singular, indicating that  for
  $\epsilon_r=\mathrm{const}$ and $\epsilon_{g,a,r}\to0$ the fixed point
  near $\Phi^{\mathrm{(a)}}\approx2\pi$ disappears. To ensure $\epsilon\ll1$ a regularizing
  expansion is needed. For instance, we may introduce the parameter
  $\delta$ such that
  $\epsilon_a=a\delta$, $\epsilon_g=g\delta$,
  $\epsilon_r=\rho\delta^2$; $0<a,g,\rho\ll1/\delta$. With
this expansion we obtain:
\begin{equation}
  \label{eq:6}
  \epsilon = \frac{-2\pi\rho}{a+g}\delta + \mathrm{O}(\delta^2),
\end{equation}
which, in turn, gives us:
\begin{equation}
  \label{eq:5}
|F'|=1-(a+g)\delta + \mathrm{O}(\delta^2),
\end{equation}
that is, because $a,g>0$, the CML regime is always stable in
  the lowest order (assuming it exists
  and $\epsilon\ll1$). In particular, this point is becoming the only attracting
  one if the nonlasing state is unstable, thus giving rise to CML
  directly at threshold.
Although the stability of the corresponding cycle does not
automatically ensures the stability of the fundamental mode-locking
regime
because instability at nonzero wavenumber can
  arise, we can often take the cavity short enough to suppress such
  instabilities as was shown in
  \cite{arkhipov2014selfa}. 

In conclusion, we developed a diagram technique, allowing to
study the CML
regimes qualitatively, and demonstrated existence of stable
limit cycles in the system, corresponding to CML
regimes. Moreover, our results show that such regimes
  can appear directly at the laser threshold. Although the area theorem is
applicable for the case of large $T_2$, practically ``large'' means
only that $T_2\gg$ the round-trip time. Our numerical simulations
show also that the predictions of the theory are valid even if
these times are comparable (but still $T_2$ is not too small).
We also extended the CML to the case when $m_d$, the ratio of the
  transition dipole moments of the absorber and the gain
  media is arbitrary.
In contrast to previously considered $2\pi$ configurations,
  in this case,
   $0\pi$ pulses as well as $2n\pi$ pulses can be
  obtained. The later  can be called super-CML because more than one
  Rabi oscillation per cavity round-trip appears. Such $2n\pi$
  pulses are however unstable and split 
  into $n$ sub-pulses, each of them having area $2\pi$.
  Although CML has not been realized experimentally yet, the most of
  the basic aspects of SIT including the area theorem were verified
  experimentally
  \cite{mccall1969self,Allen:book,rothenberg1984observation,schuettler08}; In
  particular, coherent phenomena were recently observed in quantum dots \cite{karni2013rabi,kolarczik}. Thus, we
  hope that the present letter will facilitate further efforts on its
  experimental realization.

\section*{Funding Information}
This work was partially financially supported by Government of Russian Federation, Grant 074-U01.



\end{document}